\documentclass[prd,showkeys,floatfix,nofootinbib,
               preprint,12pt,
               tightenlines,fleqn]{revtex4-1} 

\usepackage{amsmath,amssymb,revsymb,graphicx,dcolumn}
\usepackage{leftidx}

\usepackage{marginnote}

\newcommand{\beq}{\begin{equation}}
\newcommand{\eeq}{\end{equation}}
\newcommand{\beqa}{\begin{eqnarray}}
\newcommand{\eeqa}{\end{eqnarray}}
\newcommand{\bsubeqs}{\begin{subequations}}
\newcommand{\esubeqs}{\end{subequations}}

\usepackage{etoolbox}
\makeatletter
\patchcmd{\frontmatter@RRAP@format}{(}{}{}{}
\patchcmd{\frontmatter@RRAP@format}{)}{}{}{}
\renewcommand\Dated@name{}
\makeatother

\begin{document}

\title[]
      {Free-fall non-universality in quantum theory}      
\author{Viacheslav~A.~Emelyanov}
\email{viacheslav.emelyanov@partner.kit.edu}
\affiliation{Institute for Theoretical Physics\\
Karlsruhe Institute of Technology\\
76131 Karlsruhe, Germany\\}

\begin{abstract}
\vspace*{2.5mm}\noindent
We~show by embodying the Einstein equivalence
principle -- local Poincar\'{e} invariance -- and general covariance in quantum~theory
that wave-function spreading rules out the universality of free fall,~i.e.
the free-fall trajectory of a quantum (test) particle depends on its internal properties.~We provide~a quantitative estimate of the free-fall~non-universality
in terms of the E\"{o}tv\"{o}s
parameter,~which turns out to be measurable in atom interferometry.
\end{abstract}

\maketitle

\section{Introduction}

According to Newton's gravitational law,~any body having a non-zero gravitational~mass~is a source of gravity.~It~is a consequence of numerous experiments that gravitational mass~$M_g$~of a macroscopic
body is equal with good accuracy to its inertial mass $M_i$.~So,~one~might~assume
\beqa\label{eq:gravitational-vs-inertial-classical}
\big(M_g/M_i\big)_\textrm{classical} &=& 1\,.
\eeqa
In Newton's theory,~this means that small-enough test bodies fall down equally
fast,~provided 
same initial position and velocity.~The general theory of relativity (GR) promotes~this~result
to the weak equivalence principle which is also known in the literature as the universality~of
free fall~\cite{Will}.~This principle is a core argument for modelling gravitational interaction~through
space-time geometry~\cite{Casola&etal}, where particles' trajectories correspond to geodesic
world lines.

In the framework of quantum theory, however, particles cannot be thought of as~point-like
objects which move along single world lines.~Indeed, Heisenberg's uncertainty principle~forces
to abandon the idea that position and momentum can be simultaneously defined~with~perfect
precision for quantum particles~\cite{Merzbacher}.~This quantum fuzziness originates from the fact that wave functions have finite localisation in space,~resulting in the probability of finding~a~particle~at~a
given space-time point, which is always less than unity.~This suggests that quantum particles
might not obey the weak equivalence principle,~provided its potential breach
does~not~involve
tidal gravitational forces~\cite{Will},~which do modify free-fall trajectories of extended bodies.

In this article,~we explore this conceptual conflict quantitatively.~This is achieved,~first,~by 
working in the framework of quantum field theory over curved spacetime, where the latter is
modelled by GR,~and,~second,~by implementing Einstein's equivalence principle~and~general
covariance by relating quantum fields to elementary particles.

\section{Free fall of classical particles}

According to Einstein's gravitational theory, matter is a source of a non-trivial spacetime
curvature.~The spacetime curvature is mathematically described by the Riemann tensor.~This
tensor has dimension of inverse length squared.~In other words, we can characterise the space-
time curvature by a length scale:~The bigger this length scale, the weaker a gravitation~field~is.
In particular, at the Earth's surface, it reads
\beqa
L_{\oplus} &\equiv& 
R_\oplus\,\big(R_\oplus/R_{S,\,\oplus}\big)^\frac{1}{2}
\;\approx\; 1.71{\times}10^{11}\,\text{m}\,,
\eeqa
where $R_\oplus \approx 6.37{\times}10^6\,\text{m}$ denotes the Earth's radius,~whereas
$R_{S,\,\oplus} \approx 8.87{\times}10^{-3}\,\text{m}$ stands for its Schwarzschild ($S$)
radius.~Thus,~the Earth's curvature plays a little role in the dynamics~of microscopic
objects in quantum processes taking place over time intervals much smaller than
$L_{\oplus}/c \approx 9.52\,\text{min}$, where $c \approx 2.99{\times}10^8\,\text{m}/\text{s}$
is the speed of light in vacuum.~For this reason,~we
shall neglect the Earth's curvature in what follows until~Sec.~\ref{sec:gd},~which
is also needed not~to~go beyond the application domain of the weak equivalence principle.

This approximation means that the metric tensor at the Earth's surface can be~replaced~by
the Minkowski metric $\eta \equiv \text{diag}(+1,\,-1,\,-1,\,-1)$ iff one considers local inertial~coordinates.
To this end,~we wish to introduce normal Riemann coordinates,~$y$,~defined at a given point~at
the Earth's surface, which corresponds to $y^a = 0$.~In its
vicinity,~i.e.~$|y| \ll L_\oplus$,~we~have
\beqa\label{eq:metric}
ds^2 &=& g_{ab}(y)\,dy^a dy^b
\;\approx\; \eta_{ab}\,dy^a dy^b\,,
\eeqa
where the Latin indices lie in $\{0,1,2,3\}$.~We have neglected curvature-dependent terms~on~the
right-hand side of~\eqref{eq:metric}, because of the weakness of the Earth's gravitational field.~These~terms
can be found in~\cite{Petrov}. The very fact that the metric tensor can always be locally brought~to~the
Minkowski-metric form is a result of Einstein's equivalence principle -- locally and at any~non-singular
point of the Universe, the special-relativity physics
applies~\cite{Casola&etal}.

The general principle of relativity, saying that dynamical laws of nature are the same~in~all
reference frames,~ensures that physics does not depend on coordinates utilised.~Nevertheless,
the same physical process can look different in different coordinate frames.~In particular, the local
inertial coordinates $y$ and general coordinates $x \equiv (ct,x,y,z)$ are related as follows~\cite{Petrov}:
\beqa\label{eq:diffeomorphism}
x^c &\approx& y^c - \frac{1}{2}\,\Gamma_{ab}^c\,y^ay^b\,,
\eeqa
where $\Gamma_{ab}^c$ are Christoffel symbols computed at the Earth's surface
and we have omitted~terms which depend on higher-order derivatives of metric, in accord
with the Minkwoski-spacetime approximation~\eqref{eq:metric}.~Taking into account that
the Earth's gravitational field is approximately described by the Schwarzschild geometry, we obtain
\bsubeqs\label{eq:christoffel}
\beqa\label{eq:diffeomorphism-0}
\Gamma_{ab}^0\,y^ay^b &\approx& +\frac{2g_\oplus}{c^2}\,y^3y^0\,,
\\[1mm]
\Gamma_{ab}^1\,y^ay^b &\approx& -\frac{2g_\oplus}{c^2}\,y^3y^1\,,
\\[1mm]
\Gamma_{ab}^2\,y^ay^b &\approx& -\frac{2g_\oplus}{c^2}\,y^3y^2\,,
\\[1mm]\label{eq:diffeomorphism-3}
\Gamma_{ab}^3\,y^ay^b &\approx& +\frac{g_\oplus}{c^2}\,\big((y^0)^2+(y^1)^2+(y^2)^2-(y^3)^2\big)\,,
\eeqa
\esubeqs
where the free-fall acceleration points down in the negative
$z$-direction with the magnitude~at the Earth's surface reading
\beqa
g_\oplus &\equiv& \frac{c^2R_{S,\,\oplus}}{2(R_\oplus)^2} \;\approx\; 9.81\,\text{m}/\text{s}^2\,.
\eeqa
Now, in the Riemann frame, all geodesics passing through $y^a = 0$ are straight world lines~\cite{Petrov}.
This is basically the condition which determines normal Riemann coordinates.~Considering~a
classical (point-like) particle being initially at rest in the Riemann frame, we have
\beqa\label{eq:straight-line}
y^a(\tau) &=& c\tau\delta_0^a\,,
\eeqa
where $\tau$ is the proper time and $\delta \equiv \text{diag}(+1,\,+1,\,+1,\,+1)$ is the Kronecker
delta.~It turns~into
\beqa\label{eq:curved-line}
x^a(\tau) &\approx& c\tau\delta_0^a - \frac{1}{2}\,g_\oplus\tau^2\delta_3^a
\eeqa
in the non-inertial frame associated with the Earth's surface,~where we
have~substituted~\eqref{eq:straight-line} into~\eqref{eq:diffeomorphism} and~\eqref{eq:christoffel}
to get~\eqref{eq:curved-line}.~This is the well-known result of Newton's gravitational theory,~that
explicitly demonstrates the universality of free fall in classical theory.

\section{Free fall of quantum particles}

Quantum field theory (QFT) is a mathematical formalism which enables us to successfully
describe high-energy processes taking place between particles.~This formalism is based on~the
unification of the underlying principles of quantum mechanics (QM) and the special theory~of
relativity~(SR). The observable Universe cannot be described by Minkowski~spacetime, which
is a basic mathematical structure of SR. Hence, the application of QFT~in~theoretical~particle
physics relies on the Minkowski-spacetime approximation~\eqref{eq:metric}.

It is apparent that we need to go beyond this approximation in order to describe quantum
particles in the presence of a gravitational field. We thereby wish to demand that the~Einstein
equivalence principle and the general principle of relativity be also implemented in quantum
theory.~The former principle implies then that quantum particles must locally be modelled by
wave functions which, in local inertial frames, are given by plane-wave superpositions.~In~fact,
it ensures that such quantum particles move along straight world lines in local inertial frames.
The latter principle says in turn that wave functions must transform as tensors under general 
coordinate transformations.~In particular,~a spin-zero-particle wave function~must~correspond
to a rank-zero tensor -- scalar. This ensures that the semi-classical~Einstein~field~equation~is~in
accord with general covariance.

Quantum fields are operator-valued distributions which form a
quantum-field algebra~\cite{Haag}.
To model a quantum particle in this framework,~we need to select the
operator~$\hat{a}^\dagger(\psi)$~from~this algebra,~that gives the state
$|\psi\rangle = \hat{a}^\dagger(\psi)|\Omega\rangle$ describing this particle,~where
$|\Omega\rangle$ is the quantum vacuum.~In a local Minkowski frame,~$|\psi\rangle$
must reduce to an asymptotically free state entering the definition of $S$-matrix elements
in particle physics.~To guarantee that,~we define
\beqa\label{eq:psi-creation-operator}
\hat{a}^\dagger(\psi) &\equiv& - i{\int_\Sigma}d\Sigma^a(x)
\Big(\psi(x)\partial_a \hat{\Phi}^\dagger(x) - \hat{\Phi}^\dagger(x)\partial_a\psi(x)\Big)\,,
\eeqa
where $\Sigma$ is a Cauchy surface and $\hat{\Phi}(x)$ denotes a scalar field,~because then
we~locally~recover the Lehmann-Symanzik-Zimmermann reduction
formula for the scalar field.~In general,~this formula relates $S$-matrix elements with
time-ordered products of quantum fields~\cite{Lehmann&Symanzik&Zimmermann,Srednicki},~or,~in
other words, it relates the mathematical formalism of QFT to physics.

Now,~$\psi(x)$ in~\eqref{eq:psi-creation-operator} corresponds to a wave
function,~at least in the weak-gravity limit,~i.e.~we assume that the characteristic
linear
size of $\psi(x)$ in space,~$L$,~is much smaller~than~$L_\oplus$.~In~the
Riemann frame,~Einstein's equivalence principle tells us that the wave function~of~a~spin-zero 
particle of mass $M > 0$ is a superposition of (positive-energy) plane
waves,~namely
\bsubeqs\label{eq:wave-function}
\beqa
\psi(y) &=& \frac{1}{(2\pi\hbar)^3}{\int}\frac{d^3\mathbf{K}}{2E_\mathbf{K}}\,
F_\mathbf{P}(\mathbf{K})\exp\Big({-}\frac{iK{\cdot}y}{\hbar}\Big)\,,
\eeqa
where $\hbar \approx 1.05{\times}10^{-34}\,\text{J}{\cdot}\text{s}$ is the reduced Planck constant and
\beqa
K &\equiv& (E_\mathbf{K}/c,\mathbf{K}) \;\equiv \; \big(\sqrt{(Mc)^2+\mathbf{K}^2},\mathbf{K}\big)\,,
\\[1mm]
K{\cdot}y &\equiv& \eta_{ab}K^ay^b\,.
\eeqa
\esubeqs
The function $F_\mathbf{P}(\mathbf{K})$ must have a narrow peak at
$\mathbf{K} \sim \mathbf{P}$, where $P \equiv (E_\mathbf{P}/c,\mathbf{P})$
is an initial $4$-momentum of the particle.~This is in effect required for~$\psi(y)$ to be a
localised-in-space~packet. Furthermore, the general principle of relativity forces
us to deal with~$F_\mathbf{P}(\mathbf{K}) = F(K{\cdot}P)$.~For
instance, a covariant Gaussian wave function~\cite{Naumov&Naumov,Naumov} is characterised by
\beqa\label{eq:momentum-variance}
F_\mathbf{P}(\mathbf{K}) &\propto& \exp\Big({-}\frac{K{\cdot}P}{2D^2}\Big)\,,
\eeqa
where $D > 0$ stands for momentum variance.~The covariance principle
leads,~thereby,~to~$\psi(y)$ which~is
invariant under the (local) Lorentz transformations.~This,~in particular,~ensures
that $\psi(y)$ gains a phase shift in quantum-interference
experiments~\cite{Colella&Overhauser&Werner,Bonse&Wroblewski,Asenbaum&etal-2017,Overstreet&etal},
which~is~in~agreement with the observations~\cite{Emelyanov-2020,Emelyanov-2021}.

According to Born's statistical interpretation,~the wave packet $\psi(y)$ yields the probability
amplitude of measuring the particle at a given place~\cite{Merzbacher}.~Thus,~the probability
to~find~it~some- where in space must be unity:
\beqa\label{eq:normalisation-qm}
{\int} d^3\mathbf{y}\, \psi^*(y)\,\psi(y) &=& 1\,.
\eeqa
This is a normalisation condition for the wave function $\psi(y)$ in QM.~It is evident though~that
this normalisation condition is at odds with special covariance, since the integration measure
in~\eqref{eq:normalisation-qm}~is variant under the (local) Lorentz transformations.~Therefore,~it
must be replaced~in QFT by the Klein-Gordon product of $\psi(y)$ with itself,~cf.~\cite{Laemmerzahl-1995}:
\beqa\label{eq:normalisation-qft}
i{\int} d^3\mathbf{y}\,\Big(\psi^*(y)\,\partial_0\psi(y) - 
\psi(y)\,\partial_0\psi^*(y)\Big) &=& 1\,.
\eeqa
In fact,~this equation corresponds to $\langle \psi | \psi \rangle = 1$.\,This directly
follows from the~definition~of~$| \psi \rangle$
and the canonical commutation relation of $\hat{\Phi}(x)$ and its canonical conjugate.~Note
that~\eqref{eq:normalisation-qft}~is independent on (local) inertial frames,~provided
$\psi(y)$~is~a~scalar.~This physically implies that
quantum particles are reference-frame-independent~objects,~i.e.~their very existence~does
not depend on coordinates utilised~\cite{Emelyanov-2020,Emelyanov-2021}.

We wish to derive a free-fall trajectory of the quantum
particle.~In real-world experiments,
quantum-particle trajectories are determined with the help of detectors.~Any~detector~has~a 
finite extent in space (and time).~One can describe this by the scalar $W(y)$ which is essentially
unity inside a particle detector and tends to zero outside that.~This is a window-function-like
scalar which can be understood in QFT as being due to the spontaneous breakdown~of~spatial
translation symmetry.~So,~$W(y)$ is an order parameter.~The device clicks if the particle passes
through it,~omitting details of how those interact with each other for the sake of
simplicity.~In
this case,~the particle position matches the detector location at the time moment of click.~The
covariant probability~to~find the particle~in an infinitesimal spatial volume at $y$ follows~from
\eqref{eq:psi-creation-operator} and reads
\beqa
dP(y) &\equiv& -id\Sigma^a(y)\Big(\psi(y)\partial_a \psi^*(y) - \psi^*(y)\partial_a\psi(y)\Big)\,.
\eeqa
It is non-negative and drops substantially to zero away from the wave-function support.\,Thus,
\beqa
P_W(\Sigma) &\equiv& {\int_{\Sigma}}dP(y)\,W(y)
\eeqa
gives the probability to observe the particle by this device.~Note,~$W(y)$ is~to~covariantly~limit
the integration volume to that which the detector occupies,~cf.~\cite{Pang&etal}.~If there is an array~of~such
small-enough detectors, then~the particle position can be determined with some accuracy.~On
the other hand, we have from probability theory that
\beqa
\langle y^a(\Sigma) \rangle  &\equiv& {\int_{\Sigma}}dP(y)\,y^a
\eeqa
gives the expected value of $y^a$,~which,~in physics,~corresponds to the center-of-mass position~of
the wave function $\psi(y)$.~In terms of this quantity,~the device clicks if~the wave-function~center
of mass is localised within the support of $W(y)$.

These observations suggest that the quantum-particle position corresponds~to
\beqa\label{eq:position-qft}
\langle y^a(\tau) \rangle &\equiv& i{\int_{\tau}} d^3\mathbf{y}\,y^a
\Big(\psi^*(y)\,\partial_0\psi(y) - 
\psi(y)\,\partial_0\psi^*(y)\Big)\,,
\eeqa
which turns into the quantum-mechanics definition of position expectation value in the non-relativistic
limit~$ |\mathbf{P}| \ll Mc$~\cite{Emelyanov-2021}.~Note that the position
expectation value $\langle y^c(\tau) \rangle$ depends~on the proper time $\tau$.~This is a physical
hypothesis, meaning that quantum~particles~measure~$\tau$. This, however, can be justified by
recalling that a lifetime of cosmic-ray~(relativistic)~muons~is bigger than that of muons at
rest.~This discrepancy arises due to the~time-dilation~effect~in~SR\\
\cite{Rossi&Hall}: The laboratory lifetime of the cosmic-ray muons is by a Lorentz factor
bigger~than~their proper lifetime.~This experimental result validates our hypothesis.

Consequently,~we obtain~from
\eqref{eq:wave-function},
\eqref{eq:momentum-variance}, \eqref{eq:normalisation-qft} and \eqref{eq:position-qft}~for
the spin-zero quantum particle being initially at rest ($|\mathbf{P}| = 0$)~that
\beqa
\langle y^a(\tau) \rangle &=& c\tau\delta_0^a
\eeqa
in the Riemann or, in other words, local inertial frame, while, by bearing in mind~\eqref{eq:diffeomorphism}
and~\eqref{eq:christoffel},
\beqa\label{eq:curved-line-qf}
\langle x^a(\tau) \rangle &\approx& c\tau\delta_0^a -
\frac{1}{2}\,g_\oplus{\left(\left(1 + \frac{D^2}{(Mc)^2}\right)\tau^2+\frac{\hbar^2}{4(Dc)^2}\right)}\delta_3^a
\eeqa
in the non-inertial frame associated with the Earth's surface.~The quantum result
\eqref{eq:curved-line-qf}~differs from the classical one~\eqref{eq:curved-line}
by terms to depend on internal
quantum-particle~properties.~Note,
the deviation from the geodesic depends on the characteristic quantum-particle~extent $\hbar/D$,
following from Heisenberg's uncertainty relation,~but is not due to tidal gravitational~forces.
In fact,~the tidal-force impact on free fall diminishes with decreasing extent~of a freely~falling
body,~unlike the \emph{time-dependent} correction to~\eqref{eq:curved-line}
in~\eqref{eq:curved-line-qf}.~Our result~\eqref{eq:curved-line-qf} means thus that~the weak equivalence
principle does not hold in quantum
theory~\cite{Emelyanov-2020,Emelyanov-2021}.

The origin of the free-fall non-universality in quantum theory is wave-function spreading.
Indeed,~this universal phenomenon follows from the circumstance that the wave function~$\psi(y)$
obeys the Heisenberg uncertainty principle.~This manifests itself through
\beqa\label{eq:wps}
\langle y^iy^j (\tau)\rangle &\approx&
\left(\frac{\hbar^2}{4D^2}+{\tau^2\,\frac{D^2}{M^2}}\right)\delta^{ij}\,,
\eeqa
where $i,j \in \{1,2,3\}$,
meaning that $\psi(y)$ expands in space.~The combination of this quantum-mechanical result~with
\eqref{eq:diffeomorphism-3} explains the quantum corrections to~\eqref{eq:curved-line}
in~\eqref{eq:curved-line-qf}.

Our result~\eqref{eq:curved-line-qf} may be interpreted in Newton's gravitational theory as
\eqref{eq:gravitational-vs-inertial-classical} cannot hold~in quantum theory, namely we instead have
\beqa\label{eq:gravitational-vs-inertial-quantum}
\big(M_g/M_i\big)_\textrm{quantum} &\approx& 1 + \frac{D^2}{(Mc)^2}\,,
\eeqa
because,~owing
to the time-dependent term in~\eqref{eq:wps},~it follows from~\eqref{eq:curved-line-qf} that
\beqa\label{eq:curved-line-qf-acceleration}
\frac{d^2}{d\tau^2}\,\langle z(\tau) \rangle &\approx& -g_\oplus\,\big(M_g/M_i\big)_\textrm{quantum}\,.
\eeqa
It is worth pointing out that~\eqref{eq:gravitational-vs-inertial-quantum} is a relativistic result,~because
the quantum
correction~to~\eqref{eq:gravitational-vs-inertial-classical} disappears in the
quantum-mechanics limit,~in accordance with~\cite{Laemmerzahl}.~It originates from~going
beyond Newton's theory by taking into account gravitational-length contraction,~as~this~gives 
rise to terms in~\eqref{eq:christoffel},~depending quadratically on $y^i$.~We~intend 
next to study whether~\eqref{eq:curved-line-qf}~is~at least approximately consistent
with other observables.

\section{Four-momentum of quantum particles}

The stress-energy-tensor operator for the Klein-Gordon quantum field $\hat{\Phi}(y)$
reads
\beqa
\hat{T}_{ab}(y) &=& \partial_a\hat{\Phi}(y)\partial_b\hat{\Phi}(y) - \frac{1}{2}\,\eta_{ab}
\big(\partial_c\hat{\Phi}(y)\partial^c\hat{\Phi}(y) - (Mc/\hbar)^2\hat{\Phi}^2(y)\big)\,,
\eeqa
Making use of the canonical commutation relation 
for $\hat{\Phi}(y)$~and~its canonical
conjugate $\hat{\Pi}(y)$, we obtain for the single-particle state $|\psi\rangle$ that
\beqa
\langle\psi | \hat{T}_{ab}(y) | \psi \rangle &=& 
\langle\Omega | \hat{T}_{ab}(y) | \Omega \rangle + T^{ab}(\psi(y))\,,
\eeqa
where $\langle\Omega | \hat{T}_{ab}(y) | \Omega \rangle$ stands for the quantum-vacuum
stress tensor~\cite{Pauli,Zeldovich,Weinberg} and
\beqa
T_{ab}(\psi(y)) &\equiv& 2\partial_{(a}\psi^*(y)\,\partial_{b)}\psi(y)
-\eta_{ab}\big(|\partial \psi(y)|^2-(Mc/\hbar)^2|\psi(y)|^2\big)\,.
\eeqa 
Apparently,~the quantum vacuum $| \Omega \rangle$ does not carry information about the
quantum particle
modelled by $|\psi\rangle$.~That is a no-particle state by its very definition.~This~means that we need~to
renormalise $\langle\psi | \hat{T}_{ab}(y) | \psi \rangle$ by subtracting
$\langle\Omega | \hat{T}_{ab}(y) | \Omega \rangle$ from it.\,This gives rise
to~$\langle\psi |{:}\hat{T}_{ab}(y){:}| \psi \rangle$,
where the colons mean the normal ordering,~being equal to $T_{ab}(\psi(y))$.

Taking into account that $T_{ab}(\psi(y))$ is a tensor,~we find in the frame
resting~on~the~Earth's surface~for~the particle with the initial momentum $|\mathbf{P}| = 0$ that
\beqa\nonumber
\langle p^a(\tau) \rangle &\equiv&
{\int_{\tau}} d\Sigma^c(y)\,\frac{\partial x^a}{\partial y^b}\,T_c^{b}(\psi(y))
\\[1mm]\label{eq:four-momentum}
&\approx&
Mc\left(1 + \frac{3}{2}\frac{D^2}{(Mc)^2}\right)\delta_0^a
- M g_\oplus\tau\left(1 + \frac{5}{2}\frac{D^2}{(Mc)^2}\right)\delta_3^a\,.
\eeqa
This result can be immediately obtained from $M_i\langle \dot{x}^a(\tau) \rangle$
with~\eqref{eq:curved-line-qf}, which, in classical theory, gives
particle's 4-momentum,~where the inertial mass $M_i$ has been defined via the Lagrangian
mass $M$ at the leading order of the approximation as follows:
\beqa\label{eq:inertial-mass}
M_i &\equiv& M\left(1 + \frac{3}{2}\frac{D^2}{(Mc)^2}\right).
\eeqa

These computations give an independent support for the
result~\eqref{eq:curved-line-qf-acceleration},~as we have from~\eqref{eq:four-momentum}
with~\eqref{eq:inertial-mass} that
\beqa
\frac{d}{d\tau}\,\frac{\langle p^z(\tau) \rangle}{M_i} &\approx& -g_\oplus\,\big(M_g/M_i\big)_\textrm{quantum}.
\eeqa
It should be mentioned that this derivation makes no use of Born's statistical interpretation
we have utilised above to link the quantum-particle trajectory with the wave-function~center-
of-mass position.

\section{Geodesic deviation for quantum particles}
\label{sec:gd}

The free-fall acceleration is a non-inertial-frame effect which is, accordingly, absent in~local
inertial frames.~In contrast,~the spacetime curvature is non-vanishing in all reference frames.
In particular,~it shows itself as a relative acceleration between geodesics.~This~starts~to~play~an
important role in satellite-borne experiments.

Considering a detector at rest at the origin of a Riemann frame
parametrised by $\chi$,~we~find in terms~of the normal Riemann coordinates $y$ that
\beqa
\chi^c &\approx& X^c + y^c - \frac{1}{3}\,R_{\;\;adb}^c\,\big(y^aX^dy^b - X^ay^dX^b\big)
\eeqa
where $y^a = 0$ corresponds to $X^a$ in the satellite's rest frame, and
$R_{\;\;adb}^c$ is the Riemann tensor at that point.~Taking into account that
$\langle y^c(\tau) \rangle$ gets no contribution linearly depending~on~the curvature
tensor in vacuum,~we find
\beqa\label{eq:qf-tidal-acceleration}
\frac{d^2}{d\tau^2}\,\langle \chi^c(\tau) \rangle &\approx& 
- \frac{2}{3}\,R_{\;\;adb}^c\,U^aX^dU^b \left(1 + \frac{D^2}{(Mc)^2}\right),
\eeqa
where $U^a \equiv P^a/M$ is the initial 4-velocity of the quantum particle.

This result is in accord with the geodesic deviation equation up to the factor depending~on
the internal quantum-particle properties.\,This factor fully agrees with that
in~\eqref{eq:curved-line-qf-acceleration},~suggesting
that the (passive) gravitational mass of the quantum particle is by
that factor bigger
than~its inertial mass.~This is,~apparently,~in agreement with~\eqref{eq:gravitational-vs-inertial-quantum}.

\section{Quantitative estimate}

The result~\eqref{eq:curved-line-qf-acceleration} implies quantum particles fall down faster than
classical ones.~This~effect
is negligibly small for macroscopic objects.~In particular,~one gram of iron~has~the size of~about
$6.24{\times}10^{-3}\,\text{m}$,~which may be equated to~$\hbar/D$, according to Heisenberg's
uncertainty~relation, giving $D/Mc \approx 5.63{\times}10^{-38}$.~However,~a~rubidium~atom,~${}^{85}\text{Rb}$, has the radius of $220{\times}10^{-12}\,\text{m}$
and, thus, we get the estimate~$D/Mc \approx 5.63{\times}10^{-9}$.

A dimensionless parameter,~which quantifies~relative free-fall acceleration of a pair~of~test
bodies of different composition, is known as the E\"{o}tv\"{o}s parameter~$\eta$.~We find
from~\eqref{eq:curved-line-qf-acceleration}~that
\beqa
\eta(A,B) &\approx& \frac{D_A^2}{(M_Ac)^2} - \frac{D_B^2}{(M_Bc)^2}\,.
\eeqa
It approximately reads $3.16{\times}10^{-17}$ in case of
${}^{85}\text{Rb}$~and~a~heavier~atom.~This is by five~orders~of magnitude 
smaller than the atom-interferometer sensitivity recently achieved in~\cite{Asenbaum&etal}~(see~also 
\cite{Schlippert&etal,Albers&etal}) by experimental tests of the
universality of free fall,~where the heavier atom was~the
\begin{table}[h!]
\centering
\begin{tabular}{l l | l c l | l c}
& & & $(D/Mc)^2$ &&& $(L/R_\oplus)^2$ \\
\hline
One gram of iron &&& $3.17{\times}10^{-75}$ &&& $9.59{\times}10^{-19}$ \\
Rubidium atom (${}^{85}$Rb) &&& $3.16{\times}10^{-17}$ &&& $4.77{\times}10^{-33}$\\
Potassium atom (${}^{39}$K) &&& $1.78{\times}10^{-16}$ &&& $1.02{\times}10^{-33}$\\
Hydrogen atom (H) &&& $1.14{\times}10^{-11}$ &&& $2.37{\times}10^{-35}$
\end{tabular}
\caption{The first column shows that the tinier~a~quantum~particle is,~the~bigger the effect of~wave-
function spreading influences the particle's free-fall trajectory.~In particular,~one might expect that
the effect is suppressed for Bose-Einstein condensates in free fall,~since these have~a~relatively slowly
expanding wave function,~see~\cite{Muentinga&etal,Lachmann&etal}.\,The second column
illustrates the effect~of tidal~gravitational forces on free-fall trajectories
of extended objects,~estimated within Newton's theory (see also~\cite{Emelyanov-2021}).}
\end{table}
rubidium
isotope ${}^{87}\text{Rb}$.~Yet,~the E\"{o}tv\"{o}s parameter increases by use of
lighter atoms:

Satellite-borne experiments have much better sensitivity with respect to the Earth-based
ones by quantum tests of the free-fall universality -- at the $10^{-17}$~level or better,~--~where~their
main advantage consists in the fact that these tests can potentially be made over infinite~free-
fall times~\cite{Battelier&etal}.~Their sensitivity will thus be sufficient to empirically~discover~if
wave-function spreading is more fundamental than the weak equivalence principle.

\section{Concluding remarks}

Here we
have treated the free-fall propagation of spinless quantum particles from different
standpoints.~From the perspective of a particle detector being at rest~on~the~Earth's
surface, e.g.~of that installed in the Bremen Drop Tower,
a quantum particle falls down faster than~its classical counterpart:
\beqa\label{eq:result-1}
\frac{d^2}{d\tau^2}\big(\langle z(\tau) \rangle - z(\tau)\big)
&\approx& - \frac{g_\oplus D^2}{(Mc)^2}\,.
\eeqa
This effect is due to wave-function spreading and gravitational-length contraction,~both~well
known in quantum theory and general relativity,~respectively.~Moreover,~from~the~perspective
of a detector that freely falls down in the vicinity of the Earth's surface,~the quantum~particle 
approaches this detector in the horizontal direction faster than its classical counterpart,~while
moves faster away from it in the vertical one:
\beqa\label{eq:result-2}
\frac{d^2}{d\tau^2}\big(\langle \chi^c(\tau) \rangle - \chi^c(\tau)\big) &\approx& 
-\,\frac{D^2}{3(L_\oplus M)^2}\,\big(X\delta_1^c + Y\delta_2^c - 2Z\delta_3^c\big)\,,
\eeqa
as it follows from~\eqref{eq:qf-tidal-acceleration} with $X = (0,X,Y,Z)$ and $U = (c,0,0,0)$.~This is a~consequence~of~the interplay of wave-function
spreading and the Earth's curvature.

It is a result of lots of experiments that QFT over Minkowski space locally makes physical\\
sense, although the observable Universe is actually curved.~This observation~implies~that~both
Einstein's equivalence principle and general covariance must be built into~quantum theory~for
that to be in accordance with observations in particle colliders.~This line of reasoning~leads~to
our model of quantum particles in the presence of a gravitational field.~It gives~the
results~\eqref{eq:result-1} and~\eqref{eq:result-2},~which might be experimentally testable in the near future.

Still,~there are two possible outcomes of these tests.~If it will be experimentally discovered
that the free-fall trajectory of a quantum (test) particle depends on its internal properties~in
accord with our results, then the weak equivalence principle -- one of the underlying ideas~of
GR -- should be re-thought in quantum theory.~It is worth pointing out that this circumstance
does not imply any modifications of the coupling of gravity to matter fields,~since our results
are based on the gravity theory described by a single space-time geometry.~If otherwise,~the
wave-function description~of quantum particles should be refined in GR.~In~either~case,~these
will improve our insight of both quantum theory and gravity.


\begin{thebibliography}{99}

\bibitem{Will}
C.M.\,Will, Living Rev. Relativ. {\bf 17} (2014) 4.

\bibitem{Casola&etal}
E.\,Di\,Casola, S.\,Liberati, S.\,Sonego,
Am. J. Phys. {\bf 83} (2015) 39.

\bibitem{Merzbacher}
E.\,Merzbacher, 
\emph{Quantum Mechanics}
(John Wiley \& Sons, Inc., 1998).

\bibitem{Petrov}
A.Z.\,Petrov, 
\emph{Einstein Spaces}
(Pergamon Press Ltd., 1969).

\bibitem{Haag}
R.\,Haag,
\emph{Local Quantum Physics. Fields, Particles, Algebras} (Springer-Verlag, 1996).

\bibitem{Lehmann&Symanzik&Zimmermann}
H.\,Lehmann, K.\,Symanzik, W.\,Zimmermann, Nuovo Cimento {\bf 1} (1955) 205.

\bibitem{Srednicki}
M.\,Srednicki,
\emph{Quantum Field Theory}
(Cambridge UP, 2007).

\bibitem{Naumov&Naumov}
D.V.\,Naumov, V.A.\,Naumov,
J. Phys. G: Nucl. Part. Phys. {\bf 37} (2010) 105014.
 
\bibitem{Naumov}
D.V.\,Naumov,
Phys. Part. Nuclei Lett. {\bf 10} (2013) 642.

\bibitem{Colella&Overhauser&Werner}
R.\,Colella, A.W.\,Overhauser, S.A.\,Werner, Phys.~Rev.~Lett. {\bf 34} (1975) 1472.

\bibitem{Bonse&Wroblewski}
U.\,Bonse, T.\,Wroblewski, Phys.~Rev.~Lett. {\bf 51} (1983) 1401.

\bibitem{Asenbaum&etal-2017}
P.\,Asenbaum \emph{et al.}, Phys.~Rev.~Lett. {\bf 118} (2017) 183602.

\bibitem{Overstreet&etal}
C.\,Overstreet \emph{et al.}, Science {\bf 375} (2022) 226.

\bibitem{Emelyanov-2020}
V.A.\,Emelyanov, Eur. Phys. J. C {\bf 81} (2021) 189.

\bibitem{Emelyanov-2021}
V.A.\,Emelyanov, Eur. Phys. J. C {\bf 82} (2022) 318.

\bibitem{Laemmerzahl-1995}
C.\,L\"{a}mmerzahl,
Phys. Lett. A {\bf 203} (1995) 12.

\bibitem{Pang&etal}
B.H.\,Pang \emph{et al.},
Phys. Rev. Lett. {\bf 127} (2016) 090401.

\bibitem{Rossi&Hall}
B.\,Rossi, D.B.\,Hall, 
Phys. Rev. {\bf 59} (1941) 223.

\bibitem{Laemmerzahl}
C.\,L\"{a}mmerzahl,
Gen. Rel. Grav. {\bf 28} (1996) 1043.

\bibitem{Pauli}
W.E.\,Pauli,
in \emph{Exclusion principle and quantum mechanics} (Nobel lecture, 1946).

\bibitem{Zeldovich}
Ya.B.\,Zeldovich,
Sov. Phys. Usp. {\bf 11} (1968) 381.

\bibitem{Weinberg}
S.\,Weinberg,
Rev. Mod. Phys. {\bf 61} (1989) 1.

\bibitem{Asenbaum&etal}
P.\,Asenbaum \emph{et al.},
Phys. Rev. Lett. {\bf 125} (2020) 191101.

\bibitem{Schlippert&etal}
D.\,Schlippert \emph{et al.},
Phys. Rev. Lett. {\bf 112} (2014) 203002.

\bibitem{Albers&etal}
H.\,Albers \emph{et al.},
Eur. Phys. J. D {\bf 74} (2020) 145.

\bibitem{Muentinga&etal}
H.\,M\"{u}ntinga \emph{et al.},
Phys. Rev. Lett. {\bf 110} (2013) 093602.

\bibitem{Lachmann&etal}
M.D.\,Lachmann \emph{et al.},
Nat. Commun. {\bf 12} (2021) 1317.

\bibitem{Battelier&etal}
B.\,Battelier \emph{et al.}, Exp. Astron. {\bf 51} (2021) 1695.

\end{thebibliography}
\end{document}